\documentclass[prl,epsfig,aps,superscriptaddress,showpacs,twocolumn]{revtex4}
\usepackage{epsfig}
\usepackage{latexsym}
\usepackage{graphicx}
\usepackage{amssymb}

\begin{document}
\title{Fast Adaptive Flat-histogram Ensemble for Calculating Density of States and Enhanced Sampling in Large Systems}

\author{Xin Zhou 
}

\affiliation{
Asia Pacific Center for Theoretical Physics
and Department of Physics, Pohang University of Science and Technology,
Pohang, Gyeongbuk 790-784, Korea}



\author{Yi Jiang }
\affiliation{Theoretical Division, Los Alamos National Laboratory, 
Los Alamos, NM 87545, USA}

\date{\today}

\begin{abstract}
We presented an efficient algorithm, fast adaptive flat-histogram ensemble (FAFE), to estimate the density of states (DOS) and to enhance sampling in large systems. FAFE calculates the means of an arbitrary extensive variable $U$ in generalized ensembles to form 
points on the curve $\beta_{s}(U) \equiv \frac{\partial S(U)}{\partial U}$, the derivative of the logarithmic DOS. Unlike the popular Wang-Landau-like (WLL) methods, FAFE satisfies the detailed-balance condition through out the simulation and automatically generates non-uniform $(\beta_{i}, U_{i})$ data points to follow the real change rate of $\beta_{s}(U)$ in different $U$ regions and in different systems. 
Combined with a $U-$compression transformation, FAFE reduces the required simulation steps from $O(N^{3/2})$ in WLL to $O(N^{1/2})$, where $N$ is the system size. 
We demonstrate the efficiency of FAFE in Lennard-Jones liquids with several $N$ values.  
More importantly,  we show its abilities in finding and identifying different macroscopic states including meta-stable states in phase co-existing regions.

\end{abstract}

\pacs{05.10.-a, 02.70.Ns, 64.70.D-}

\maketitle 

Complex systems, such as proteins, glasses, and many materials in the phase-coexisting regions, can be characterized as having a rugged potential energy landscape.
The molecular simulations of such systems face one central problem: how to generate samples that cover enough of the energy landscape to characterize the properties of the systems correctly. 
 Normal Metropolis Monte Carlo and molecular dynamics techniques often only 
cover part of the conformational space because of the existence of high free-energy barriers.
Over the last decade, some enhanced sampling techniques, such as the replica exchange methods (REM)~\cite{EarlD2005} and many flat-histogram techniques~\cite{BergN1991,WangL2001b,LaioP2002,ShellDP2002,YanP2003,ZhouB2005,ZhouJKZR2006,KimSK2006},
have been developed in attempts to systematically 
cover more conformation space.  
In particular, the flat-histogram techniques prescribe a collective variable and apply modified distribution functions to flatten the visited histogram in this one-variable space. 

Here we propose the density of states (DOS) to generate the flat-histogram ensemble. The DOS is defined as 
$\Omega(U) \equiv \int dr \delta (U - U(r))$, 
where $r$ is the conformation vector and $U(r)$ can be any collective variable. 
We define 
$S(U) \equiv \ln \Omega(U)$, and $\beta_{s}(U) \equiv \frac{\partial S(U)}{\partial U}$.  While $U$ is the potential energy, $S(U)$ would correspond to the entropy and $\beta_{s}(U)$ is the inverted statistical temperature. 
While the visited histogram of simulations in a ${\tilde S}(U(r))$ ensemble, which is defined as $P(r) \propto e^{-{\tilde S}(U(r))}$
$H(U) \propto e^{S(U)-{\tilde S}(U)}$, 
is almost flat, $S(U)$ can be well estimated by $S(U) = {\tilde S}(U) + \ln H(U)$. 
We can generate a sample in the ${\tilde S}(U)$ ensemble, and calculate the mean of an arbitrary variable $Q(r)$ in an arbitrary ensemble $f(X)$ 
\begin{eqnarray}
\langle Q \rangle_{f(X)}
= \frac{\sum_{k} Q(r_{k}) w_{k}}{\sum_{k} w_{k}}, 
\label{reweight-f-average}
\end{eqnarray}
where $r_{k}$ is the $k^{th}$ conformation of the sample~\cite{weighted-fluctuation}, 
and $w_k$ is the weight factor defined as:
$w_{k} = e^{{\tilde S}(U(r_{k})) - f(X(r_{k}))}$.
The statistical error of this weighted mean is proportional to $B^{-1/2}$, where $B$ is 
the effective size of the weighted sample,
$B = (\sum w_{k})^{2} / \sum w_{k}^{2}$. 
Only when sufficient conformations of the sample are located inside the important regions of the $f(X)$ ensemble ({\it i.e.}, $B$ is large enough), will the weighted sample provide a good estimate of the ensemble mean. 

The Wang-Laudau (WL) algorithm~\cite{WangL2001b} estimates $S(U)$ by updating ${\tilde S}(U)$ from an initial guess until the visited histogram is roughly a constant. 
The WL algorithm can be easily implemented and has been widely 
applied~\cite{ZhouSTL2006,PoulainCABD2006,MastnyP2005}.
However, two features of this algorithm would limit its applicability.  First, the required number of simulation steps is at least $O(N^{2})$~\cite{ZhouB2005}, where $N$ is the size of system; hence simulations of large systems will be costly.  This estimate arise from the fact that the area below the curve $S(U)$ is order-$N^2$ (because both $S$ and $U$ are proportional to $N$), which must be filled by using the small kernel functions with presumed width and height in WL~\cite{ZhouB2005}.  Second, detailed balance is not satisfied in the WL algorithm~\cite{YanP2003}, which leads to potential numerical instabilities because of the correlations between successive conformations in the simulations.  
In a recent extension to the original WL, the statistical-temperature molecular dynamics (STMD) method~\cite{KimSK2006}, 
the intensive variable $T(U) = \beta_{s}^{-1}(U)$, instead of the extensive variable $S(U)$,  is updated at each simulation step. This method is faster than the original WL, since we do not need to fill the $N$-order height of the curve $S(U)$ due to the replacement. 
However, STMD still needs to divide the extensive variable $U$ into order-$N$ homogeneous bins with a prescribed bin size and it also does not satisfy detailed balance. Thus   
the needed simulation steps are estimated to be $O(N^{3/2})$.  Another improvement of WL, the multimicrocanonical method (MMCM)~\cite{YanP2003}, calculates the microcanonical temperature $T_{m}(E)$ from the microcanonical mean, then relates the reversed $T_{m}(E)$ to 
$\beta_{s}^{-1}(E)$ based on the thermodynamical equivalence of different temperature definitions.  The MMCM has shown better numerical stability, but is also order $N^{3/2}$ because it similarly needs an $N$-order uniform grid in $U$ before starting a simulation.  Extending MMCM to the cases where the thermodynamical equivalence can not be used ({\it e.g.},  
$U$ is not the total potential energy of system), is nontrivial. 
 

In this letter, we present a more efficient algorithm, fast adaptive flat-histogram ensemble (FAFE), 
to accurately estimate the DOS and enhance sampling in large general systems. 
FAFE calculates the ensemble means and fluctuations of $U$ using 
a series of generalized ensembles,  then adaptively forms the points $(\beta_{i}, U_{i})$ on the curve $\beta_{s}(U)$.  
These $U_{i}$ points non-uniformly distributed in $U$ space and the intervals are dependent on  both $U$ and $N$ to follow the real shape of the curve $\beta_{s}(U)$: we automatically form many $U_{i}$ points in regions where $\beta_{s}(U)$ varies quickly, and few $U_{i}$ points in the regions where $\beta_{s}(U)$ varies slowly. 
A compression transformation of $U$ is also applied to reduce the visits to the $U$ regions where $\beta_{s}(U)$ has been well estimated. Therefore the required simulations steps of FAFE to estimate $S(U)$ are not more than $N^{1/2}$.  
Due to the good numerical stability of FAFE, we can visit and identify coexistent phases (including meta-stable phases) by constructing the DOS for each phase in the system,  
where the Wang-Landau like (WLL) methods cannot work unless the exact order parameters are known.  

In a $f(U)$ ensemble, 
the histogram, $H_{f}(U) \propto e^{S(U)-f(U)}$, can be approximated as the sum of a few Gaussian functions, 
$H_{f}(U)  \propto \sum C_{i} \exp\{-\frac{1}{2 \sigma_{i}^{2}} (U-U^{i})^{2}\}$,  
in the limit of large $N$. 
If $H_{f}(U)$ is a single-peak function, 
the mean of $U$ in the $f(U)$ ensemble, $\langle U \rangle_{f}$, is approximately equal to the maximal point of $H_{f}(U)$, we have,  
\begin{eqnarray}
\beta_{s}(\langle U \rangle_{f}) &=& \frac{\partial f}{\partial U}|_{\langle U \rangle_{f}}, \nonumber \\
\frac{\partial \beta_{s}}{\partial U}|_{\langle U \rangle_{f}} 
&=& - \sigma_{f}^{-2} + \frac{\partial^{2} f}{\partial U^{2}}|_{\langle U \rangle_{f}}, 
\label{extremepoint}
\end{eqnarray}
where, $\sigma_{f}$ is the fluctuation of $U$ in the $f(U)$ ensemble. 
Thus, we can calculate $\langle U \rangle_{f}$ and $\sigma_{f}$ in different $f(U)$ ensembles to generate a series of points on the curve $\beta_{s}(U)$. The complete curve of $\beta_{s}(U)$ can be interpolated from these points, and ${\tilde S}(U)$ can be estimated by integrating $\beta_{s}(U)$. 
This approximation, eq.(\ref{extremepoint}), is sufficient in $U$ regions where simulations based on the constructed ${\tilde S}(U)$ can visit sufficiently. 

Different $f(U)$ ensembles, such as 
$f(U) =\beta U$ (canonical ensembles), and  
$e^{-f(U)}= (E-U)^{l-2 \over 2} \Theta(E-U)$ (microcanonical ensembles), can be used to construct $\beta_{s}(U)$.
It is also possible to use generalized ensembles, such as  
\begin{eqnarray}
f(U) = \beta_{0} U + \frac{1}{2} a (U-U_{0})^{2},
\label{general-ensemble}
\end{eqnarray} 
to generate points on $\beta_{s}(U)$. 
Because the derivative $\frac{ \partial \beta_{s} } { \partial U }$ is positive, phase-coexistence regions are unstable in canonical ensembles. But in a generalized ensemble, e.g.,  $a > \frac{ \partial \beta_{s}}{ \partial U }$, they   
may be stabilized.   
Thus,  once we choose suitable parameters for $(\beta_{0}, U_{0},a)$,  we can even generate the $\beta_{s}(U)$ data points in the phase-coexitence regions. 

The implementation of FAFE is rather straightforward. Without losing any generality, we choose $U$ as the total potential energy, and define the statistical temperature $T_{s}(U)=\beta_{s}^{-1}(U)$ to illustrate it:
(i) Choose an initial $T_{0}$ and set $T_{s}(U) \equiv T_{0}$, then ${\tilde S}(U)=U/T_{0}$; 
(ii) Simulate a segment of trajectory in the current ${\tilde S}(U)$ ensemble and generate an equilibrium sample; 
(iii) Choose a few $f(U)$ functions ({\it e.g.} $f(U)=U/T_{i}$ for a few different $T_{i}$ ) and calculate the ensemble means ${\bar U}_{i}$ and fluctuations $\sigma_{i}$ from the generated sample at (ii) based on the eq.(\ref{reweight-f-average}). 
{\it We choose $T_{i}$ from the current $T_{s}(U)$ function: the corresponding 
${\tilde U}_{i}$, which satisfies the equation $T_{i} =T_{s}({\tilde U}_{i})$, must be located in the current visited regions. In addition, $|U_{i}-U_{i-1}| \sim \sigma_{i-1}$, where $\sigma_{i-1}$ is the fluctuation of $U$ at the temperature $T_{i-1}$.  Moreover, we only need to apply the mean and fluctuation at $T_{i}$ when the effective size of the weighted sample, $B$, is sufficiently large. 
Note that we can calculate the means and fluctuations at many $T_{i}$ from the same sample without additional simulations. The CPU time required for calculation of the weighted mean is almost negligible compared to that for generating samples; thus we can generate more data points without much additional computational cost; }
%
(iv) From the discrete points $({\bar U}_{i}, T_{i})$ (and the derivative of $T_{s}(U)$, $T^{\prime}_{i}=T^{2}_{i} / {\sigma^{2}_{i}}$), interpolate $T_{s}(U)$~\cite{linear-interpolation} and extrapolate it linearly beyond the last $U_{i}$~\cite{extrapolation}. Integrate ${\tilde S}(U)$ from the new $T_{s}(U)$;
(v) Repeat (ii), (iii) and (iv) until the $T_{s}(U)$ and ${\tilde S}(U)$ curves are complete.   
In Step (iii), we can check for the possible multi-peak characteristics of the weighted sample by 
using the generalized ensembles described by eq.(\ref{general-ensemble}). 
For example, we first choose $a=0$ ({\it i.e. canonical ensembles}) to calculate $(T_{i},{\bar U}_{i})$, then set $\beta_{0}=T^{-1}_{i}$, $U_{0}={\bar U}_{i}$ and use a positive $a$ to calculate the means in the new $f(U)$ ensemble, comparing the results to determine if larger $a$ should be used to depress the possible multi-peak effects.  In practice, unless simulations are unexpectedly found to be trapped, this checking step is not necessary. 
The points $(\beta_{i}, U_{i})$ can be calculated by using many samples generated in different segments of simulations in (ii) rather than only using a single segment of simulations,
${\bar U} \leftarrow \frac{ \sum {\bar U}^{j} B^{j} } {\sum B^{j}}$,
where  ${\bar U}^{j}$ and $B^{j}$ are the the weighted mean of $U$ and the corresponding $B$ factor from the $j^{th}$ sample, respectively, and the statistical error of ${\bar U}$, which is measured by the total $B = \sum B^{j}$ factor, can be depressed. 

\begin{figure}
\includegraphics[width=6cm]{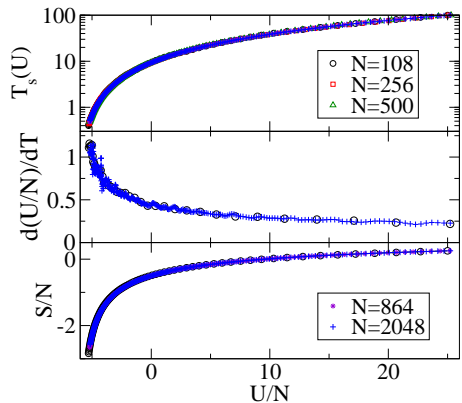}
\caption{ The $T_{s}(U)$, $c_{v} = (N T_{s}^{\prime}(U))^{-1}$ and 
$S(U)$ in systems with different size $N$ are calculated in a large temperature interval. 
 The data points are adaptively generated to form the flat-histogram ensemble with the required accuracy, so that more data points are formed in larger systems and in the lower temperature region. 
}
\label{tselj}
\end{figure}

In many cases, the $T_{s}(U)$ may be well estimated  
locally, ({\it i.e.}, the $B$ factor of the corresponding $\bar U$ is large enough), before the other 
regions have been explored.  For example, in usual physical systems, $T_{s}(U)$ at high temperature can be obtained before the low temperature region is visited. 
The visit frequency in the large-$U$ regions, which is proportional to $N$ even in the flat-histogram ensemble, can be further depressed to force the simulation to focus more on the low-$U$ region.   
To do so, we  transform $U$ to a new variable $Y$ by $\frac{d Y}{d U} = g_{c}( \frac{ U-U_{c} } {b} )$, where the compression function $g_{c}(z)$ slowly approaches zero while $z \rightarrow \infty$, but unity while $z \le 0$.  
Here $U_{c}$ and $b$ are parameters. 
One possible choice is $g_{c}(z) = \frac{ e^{-z} + 1 } { e^{-z} + z + c }$ with the parameter $c \sim 2$ or $3$. 
At the end of each segment of simulations, we reset $U_{c}$ at the lower boundary of the explored 
region and choose $b$ to make the expected visit probability above $U_{c}$ comparable to that below $U_{c}$, then use ${\tilde S}(Y)$ to generate the flat-$Y$ sample. 
Here ${\tilde S}(Y(U)) = {\tilde S}(U) - \ln g_{c}( \frac{ U-U_{c} } {b} )$, and 
the corresponding $P(U) \sim 1/(U-U_{c})$ while $U>U_{c}$. 
The flat$-Y$ simulations can still reach the high-$U$ region to be ergodic, but mainly visit the low-$U$ region to estimate $T_{s}(U)$ in the unexplored region. 
This compression transformation significantly decreases the computational cost  of our algorithm for very large systems. 

\begin{figure}
\includegraphics[width=5cm]{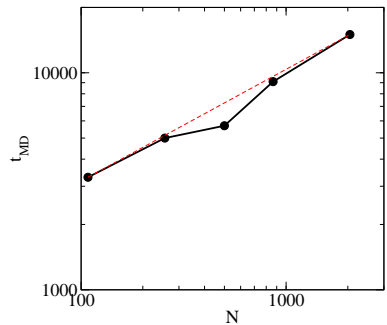}
\caption{The efficiency of our algorithm: the required MD steps grow as  $N^{1/2}$. 
}
\label{nt}
\end{figure}

To illustrate FAFE, we apply it to truncated-and-shifted Lennard-Jones (LJ) fluids 
($r_{c} = 2.5 \sigma_{lj}$) at reduced density 
$\rho = 0.8$. We choose $U$ as the total potential energy, and aim to 
calculate DOS between $T_{l}=0.5$ and $T_{h}=100$, which corresponds to a very large $U$ range in large systems.    
The ensemble mean of $U$ is sufficiently exact if the error in 
$\bar U$ is smaller than $5$ percent of the fluctuation of $U$ ({\it i.e.}, $B > 400$). 
We apply molecular dynamics (MD) simulations with a Langevin thermostat and save a conformation  
 every $10$ MD steps.  $T_{s}(U)$ is calculated and updated every 
$10^{5}$ steps, when the parameters of the compressive transformation are updated.
In our simulations, the distance between neighboring $U_{i}$ points is far larger than that in WLL~\cite{KimSK2006} in high-$T$ regions and large-$N$ systems, but automatically decreases in low $T$ regions and small $N$ systems. 
Figure (\ref{tselj}) shows that  $T_{s}(U/N)$, $c_{sv} =(d T_{s}/d(U/N))^{-1}$ and $s(u) = S(U/N)/N$ of systems 
with several different $N$ values are quickly generated in the large temperature regions.  The curves for different $N$ all collapse into single curves as expected. 
Figure (\ref{nt}) shows the efficiency of FAFE, the number of MD steps 
$t \propto N^{\gamma}$ with $\gamma \approx 0.5$. 
 
\begin{figure}
\includegraphics[width=6cm]{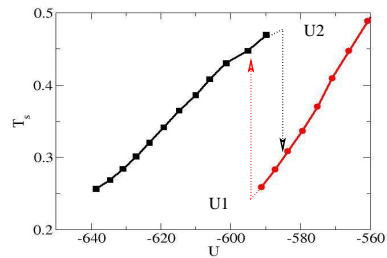}
\caption{ $T_{s}(U)$ at super-saturated liquid and solid phases in LJ system 
with $N=108$, respectively. The transition can easily happen while $U< U_{1}$ 
(from liquid to solid) and while $U> U_{2}$ (from solid to liquid). 
}
\label{pt108}
\end{figure}

As shown in Fig.(\ref{pt108}), in the lower temperature region, $T_{s}(U)$ has multiple branches that correspond to the DOS of some separated parts in conformational space. 
For a segment of simulated trajectory, we may calculate the corresponding $(T, \langle U \rangle_{T})$ points, and learn which branch of $T_{s}(U)$ the points are located on ({\it i.e.}, which macroscopic state the segment of trajectory resides in). 
The trajectories generated under the high-$U$ branch of $T_{s}(U)$ (liquid phase) have the expected flat histograms in the region $U>U_{1}$, but quickly transition into the low-$U$ region (solid phase). Similarly, simulations under the low-$U$ branch of $T_{s}(U)$ (solid phase) visit the $U<U_{2}$ region with equal probabilities, but transition into the liquid phase. The overlap of the two branches indicates that the potential energy of the liquid phase can be lower than that of the solid phase, and $U$ is definitely not the proper order parameter of the liquid/solid phase transition. It is very hard to visit (and more difficult to identify)  the low-energy liquid and high-energy solid conformations in current existing simulation techniques. Thus, FAFE, which is able to visit more parts of the configuration space and identify them by forming their own DOS without requiring the prior knowledge of precise order parameters, might provide new promise in understanding phase transition processes.  

\begin{figure}
\includegraphics[width=6cm]{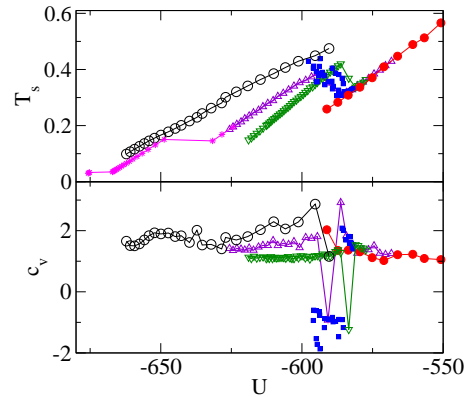}
\caption{ The $T_{s}(U)$ function in a LJ system with $N=108$ and $\rho=0.8$ at the 
low temperature region. Each branch of the function corresponds a macroscopic phase. 
The unstable liquid/solid coexistence region, where $T_{s}^{\prime}(U)=(N c_{v})^{-1}$ are negative, is shown by the blue filled squares. 
The green down-triangles corresponds to the complete face center cubic solid which is a metastable state in the current $\rho$. 
Some phase transitions from metastable solids to the stable solid (the black circles) can be found at lower temperature. 
While $T_{s}(U) \rightarrow 0$, more complex phase behaviors are expected.
}
\label{phasetransition}
\end{figure}

It is possible to calculate the total $S(U)$ and $T_{s}(U)$ from these obtained $T_{s}(U)$ branches~\cite{total-entropy}. 
However, even in the exact $S(U)$ ensemble, it is still not guaranteed that simulations can freely transition between the states. 
In this case, the visited probability of conformation inside each iso-$U$ is constant, $P(r) \propto e^{-S(U)}$, then any small-fraction part in an iso-$U$ subspace is still rarely visited although the total visited probability of the iso-$U$ subspace is not small. If a small-fraction part corresponds to the passage between two states, simulations in the $S(U)$ ensemble still do not easily pass the passage. 
The $S(U)$ ensemble only provides chances to search transition passages in more iso-$U$ subspaces. 
This is a general limitation in all the low-dimension flat-histogram ensemble methods. 
The algorithms with good stabilities ({\it e.g.}, simulations satisfying detailed balance) are helpful in analyzing and (partly) overcoming the limitation. 
For example, FAFE generates the $T_{s}(U)$ and $S(U)$ (with arbitrary constant) in each visited region instead of the expected total $T_{s}(U)$ and $S(U)$ of the whole conformational space. Thus, even when the equilibration among the regions is difficult, it is still possible to provide many useful results.  From the branches of $S(U)$ or $T_{s}(U)$, it is possible to construct the $S(U)$ for whole the conformational spac~\cite{total-entropy}. 
In addition, we can apply the generalized ensembles described by eq.(\ref{general-ensemble}) to find more (possibly metastable) states. For example, besides the stable liquid and solid phases in the LJ system, we detect several metastable solid states with different crystalline structures as well as the unstable multi-phase coexistence, which are shown 
in Fig.(\ref{phasetransition}). These complex solid phases arise from the fact that the system is fixed at 
$\rho=0.8$ so the interparticle distance in the face-center cubic structure does not match the equilibrium distance of LJ potential, $r_{0}=2^{1/6}$.  
  
The flat-histogram ensembles based on multiple collective variables are expected to be better for  enhanced sampling than single variable~\cite{ZhouSTL2006,MastnyP2005}. 
The required computational cost will be $O(N^{\gamma d})$, where $d$ is the number of the collective variables. 
FAFE ($\gamma=1/2$), which is much faster than WLL ($\gamma=1.5 \sim 3$), is efficient in sampling large complex systems, such as glasses and proteins. It is useful to construct the canonical DOS $\Omega(U;T)=\int e^{-V(r)/T} \delta(U-U(r)) dr$ instead of the normal DOS in FAFE in order to keep the connectivity of macromolecules and more focus on the desired temperature $T$, if  the total potential energy is not chosen as $U$. 
It is also possible to combine FAFE with the replica exchange method (REM): we generate the branches of $T_{s}(U)$ and use them in different replicas and exchange conformations among these replicas; thus we can use $O(N^{0})$ replicas in large systems instead of the $N^{1/2}-$order replicas required in conventional REM, which is the main limitation of applying REM. 

\begin{acknowledgments}
This work was partly supported by the US Department of Energy under contract  
No. W-7405-ENG-36. 
We are grateful to A. F. Voter for stimulating discussions. 
XZ acknowledges the Max Planck Society(MPG) and the Korea Ministry of
Education, Science and Technology(MEST) for the support of the Independent
Junior Research Group at the Asia Pacific Center for Theoretical Physics
(APCTP). 
\end{acknowledgments}








\begin{thebibliography}{16}
\bibitem{weighted-fluctuation} In the weighted sample,  the fluctuation of $Q$ is calculated as, 
$\sigma^{2} = \frac{ (\sum w_{k}) (\sum w_{k} Q_{k}^{2}) - (\sum w_{k} Q_{k})^{2}}{(\sum w_{k})^{2}}$
\bibitem{linear-interpolation} 
High-order interpolation is also possible by using  $T_{i}^{\prime}$.
\bibitem{extrapolation} In practice, we linearly extrapolate $T_{s}(U)$ in a finite range from the previous
$U_{i}$ point, and set $T_{s}(U)$ as a constant out of the range.
\bibitem{total-entropy} The total entropy $S(U)=\ln \sum_{i} \exp[S^{i}(U)]$. 
$S^{i}(U)=S^{i}_{0} + \int^{U}_{U_{i0}} T_{s}^{i}(V) dV$ and $T_{s}^{i}(U)$ are the $i^{th}$ branch of the entropy and temperature, respectively.
\end{thebibliography}

\end{document}